\documentclass[twocolumn]{article}

\usepackage{moreverb,url}
\usepackage{authblk}

\usepackage[colorlinks,bookmarksopen,bookmarksnumbered,citecolor=red,urlcolor=red]{hyperref}

\usepackage{graphicx}
\usepackage{latexsym,ifthen,rotating,calc,textcase,booktabs,color,endnotes}
\usepackage{amsfonts,amssymb,amsbsy,amsmath,amsthm}

\usepackage[text={174mm,258mm},%
papersize={210mm,297mm},%
columnsep=12pt,%
headsep=21pt,%
centering]{geometry}
\usepackage{ftnright}

\newcommand\BibTeX{{\rmfamily B\kern-.05em \textsc{i\kern-.025em b}\kern-.08em
T\kern-.1667em\lower.7ex\hbox{E}\kern-.125emX}}

\title{Lexical convergence and collective identities on Facebook}

\author[1]{Emanuele Brugnoli}
\author[1]{Matteo Cinelli}
\author[2]{Fabiana Zollo}
\author[2]{Walter Quattrociocchi}
\author[1,3]{Antonio Scala}
\date{}
\affil[1]{Applico Lab, CNR-ISC, Italy}
\affil[2]{Ca' Foscari University of Venice, Italy}
\affil[3]{LIMS, the London Institute for Mathematical Sciences, UK}
\begin{document}

\maketitle

\begin{abstract}
Recent studies, targeting Facebook, showed the tendency of users to interact with information adhering to their preferred narrative and to ignore dissenting information. Primarily driven by confirmation bias, users tend to join polarized clusters where they cooperate to reinforce a like-minded system of beliefs, thus facilitating fake news and misinformation cascades.
To gain a deeper understanding of these phenomena, in this work we analyze the lexicons used by the communities of users emerging on Facebook around verified and unverified contents.
We show how the lexical approach provides important insights about the kind of information processed by the two communities of users and about their overall sentiment.
Furthermore, by focusing on comment threads, we observe a strong positive correlation between the lexical convergence of co-commenters and their number of interactions, which in turns suggests that such a trend could be a proxy for the emergence of collective identities and polarization in opinion dynamics.
\end{abstract}

\section{Introduction}
Social media platforms have revolutionised the way people communicate, debate, access and share information. In just about a decade, they have reached billions of people around the globe, enabling exciting achievements but also posing serious challenges. In 2017 the World Economic Forum raised a warning on the potential distortion effect of social media on user perceptions of reality \cite{WEF2017}. 
Primarily driven by confirmation bias \cite{Brugnoli2019}, users online tend to join groups of like-minded people around a shared narrative, i.e. echo chambers, where they cooperate in framing and reinforcing their opinion \cite{Brown2007,Cacciatore2016,DelVicarioSR2016,DelVicarioSON2017,Kahn2004,Kumar2010,QuattrociocchiSR2014,QuattrociocchiACS2011,SunsteinBook2001}.
Confirmation bias, indeed, seems to account for users' decisions about consuming and spreading content and, at the same time, aggregation of favored information within echo chambers reinforces selective exposure \cite{Cinelli2019arXiv} and group polarization \cite{Bessi2014,Garrett2013}. Thus, echo chambers also play a role in the formation of collective identities - i.e., persons' self-identification as group members, rather than in terms of their unique, personal characteristics \cite{Boutyline2017,Spencer2007}.

To gain a deeper understanding of these phenomena, with the help of very active debunking groups, we identified all the Italian Facebook pages supporting scientific and conspiracy contents, and on a time span of five years (2010-2014) we downloaded all their public posts with the related lists of likes and comments. We do not focus on the truth value of their information but rather on the possibility to verify their claims. Indeed, the generators of scientific information and their data, methods, and outcomes are readily identifiable and available, while the origins of conspiracy theories are often unknown and their content is sharply divergent from recommended practices \cite{ZolloPONE2017}. Moreover, users usually exposed to unverified claims (e.g., conspiracy stories) are the most likely to confuse intentional false information as usual conspiracy stories \cite{Mocanu2015}. Thus we retain that the combination formed by users engaged with such contents and users who interact with pages supporting the exact opposite narrative (e.g., scientific information), represent a reasonably space of investigation for the dynamics of (mis)information spreading online. Furthermore, the engagement with science-related contents is strongly linked to political polarization \cite{Helmuth2016}. Then a better understanding of the behavior online of users active in consuming and spreading science or conspiracy contents, could give some precious insights even with regard to the mechanism of polarization in the political process.

As pointed out by previous works on the same dataset \cite{Bessi2014,BessiPONE1-2015,Bessi2015,Bessi2016,Brugnoli2019,ZolloPONE2015}, two well-separated echo chambers of users emerge around science and conspiracy narratives. Moreover, despite the very profound different nature of their contents, polarized users with opposing views consume their preferred information in a similar way in terms of volumes of both likes and comments \cite{BessiPONE1-2015}.

In this work we investigate more important analogies between the two communities with regard to the lexical behavior of their own users during written conversation.
Conversations are coordinated interactions where speakers variably lead, follow, and echo each other as they exchange ideas \cite{Carroll1980,Clark1986}. With repeated references to objects, speakers tend to reuse the same terms as they coordinate their perspectives, a phenomenon called lexical convergence \cite{Garrod1987}. This process limits and systematizes lexical variability \cite{Brennan1996} in spite of the potential for enormous variations in people's lexical choices in dialog, dubbed the vocabulary problem \cite{Furnas1984,Furnas1987}.
By reflecting conceptual coordination in dialogue \cite{Clark1986,Garrod1987}, lexical choices give important insight not only about individual processes of language use, but also about distributed ones (i.e. the use of language of a social group).

Like other kinds of social media sites, Facebook posts have options that allow members of the site to respond to them by creating a comment. As an interactive context, Facebook comment threads are polylogal, where a comment can respond either to a post or to another comment in a thread. Namely, comments represent the way in which collective debates take form around the topic promoted by posts, and from which collective identities can emerge.
Far from promoting shared interests between the co-commenters, Facebook pages could be co-opted as contexts in which rapport-threatening disagreements take place. The collective identities utilized by users who comment on these pages are therefore an important resource that might reinforce a single, shared social identity that is in line with the stated focus of the page, or that might differentiate one commenter from another. 
In the mediated context of interaction represented by the comment thread, the options for representing the commenters' collective identities are construed through the lexico-grammatical choices made by users when they are writing a comment \cite{Page2018}.

Here we first compare the lexical choices of the two echo chambers emerging around science and conspiracy narratives, both at individual and collective levels. Then, by focusing on the comment thread of a post, we test whether more conversation (intended as co-commenting activity) leads to lexical convergence. Namely, we investigate whether the lexical choices of co-commenters represent a proxy for the emergence of polarization and collective identities.

In the analysis, we first characterize the polarization of the users with respect to conspiracy or science contents by accounting for their liking activity, and we show that users are clearly split into two groups with opposing polarization. Moreover, users concentrate their commenting activity almost entirely on posts from pages of their own community.
Despite such high level of segregation, we show that the two communities of users adopt very similar vocabularies and use the same word with almost the same frequency, both at the individual and the collective level. Nevertheless, the minority of words exhibiting significant differences in frequency occurrence from one community to another, provide important insights about the kind of information processed by the two communities of users and about the overall sentiment expressed in their comments.

A static comparison of the lexical convergence between co-commenters as a function of their number of interactions, suggests a positive correlation between these two variables. Furthermore, the analysis of how lexical convergence evolves through time suggests that this trend could be a proxy for the emergence of polarization and collective identities.

\section{Materials and methods}\label{sec:mm}
\subsection{Ethics Statement}
The entire data collection process has been carried out exclusively through the Facebook Graph API, which is publicly available, and for the analysis (according to the specification settings of the API) we used only public available data (users with privacy restrictions are not included in the dataset). The pages from which we download data are public Facebook entities (can be accessed by anyone). User content contributing to such pages is also public unless the user's privacy settings specify otherwise and in that case it is not available to us.

\subsection{Data collection}
To our aim, we identified two main categories of pages: conspiracy - i.e. pages promoting contents neglected by main stream media - and science. The space of our investigation has been defined with the help of Facebook groups very active in debunking conspiracy theses (\textit{Protesi di Complotto, Che vuol dire reale, La menzogna diventa verit\`{a} e passa alla storia}). We categorized pages according to their contents and their self description.

Concerning conspiracy pages, their self description is often claiming the mission to inform people about topics neglected by main stream media. Pages like \textit{Scienza di Confine, Lo Sai} or \textit{CoscienzaSveglia} promote heterogeneous contents ranging from aliens, chemtrails, geocentrism, up to the causal relation between vaccinations and homosexuality. Conversely, science pages - e.g \textit{Scientificast, Italia unita per la scienza} are active in diffusing posts about the most recent scientific advances.
Notice that we do not focus on the truth value of their information but rather on the possibility to verify their claims. Indeed, the main difference between the two is content verifiability. The generators of scientific information and their data, methods, and outcomes are readily identifiable and available. The origins of conspiracy theories are often unknown and their content is sharply divergent from recommended practices \cite{ZolloPONE2017}.
The selection of the source has been iterated several times and verified by all the authors. To our knowledge, the final dataset is the complete set of all scientific and conspiracist information sources active in the Italian Facebook scenario. Notice that the dataset used in the analysis is the same used in \cite{Bessi2014,ZolloPONE2015,BessiPONE1-2015,Bessi2015,Bessi2016,Brugnoli2019}.

The pages from which we downloaded data are public Facebook entities (can be accessed by virtually anyone). The resulting dataset is composed of 73 public pages for which we downloaded all the posts and all the likes and comments from the posts over a time span of five years (Jan 2010 - Dec 2014). The exact breakdown of the data is presented in Table~\ref{table1}. See Supplemental material for the list of scientific and conspiracy pages, respectively.

\begin{table}[!ht]
\small\centering
\caption{
Breakdown of Facebook dataset.\label{table1}}
\begin{tabular}{llll}
\toprule
& Total & Science & Conspiracy\\
\midrule
Pages & 73 & 34 & 39\\
Posts & 271,296 & 62,705 & 208,591\\
Likes & 9,164,781 & 2,505,399 & 6,659,382\\
Comments & 1,017,509 & 180,918 & 836,591\\
Likers & 1,196,404 & 332,357 & 864,047\\
Commenters & 279,972 & 53,438 & 226,534\\
\bottomrule
\end{tabular}
\end{table}

In the analysis, we account for user interaction with respect to public posts - i.e., likes and comments. The two actions have a different meaning from the user viewpoint \cite{Viswanath2009}. A like stands for a positive feedback to the post, whereas a comment is the way in which online collective debates take form around the topic promoted by posts. Comments may contain negative or positive feedbacks with respect to the post.

\subsection{User polarization and engagement}\label{ssec:user_pol}
Let $\mathcal{P}$ be the set of all the posts and let $V$ be the set of all the users in our collection. Moreover, let $V^L$ be the set of users of $V$ who have liked at least one post of $\mathcal{P}$. Following previous works \cite{Bessi2014, BessiPONE1-2015,Bessi2016,Mocanu2015}, we define the polarization of users - i.e., the tendency of users to interact with only a single type of information - towards science and conspiracy through a simple thresholding algorithm accounting for the percentage of likes on one or the other category. Formally, we define the polarization of a user $u\in V^L$ as $\sigma_{u}=2\rho_{u}-1$, where $0\leq\rho_{u}\leq 1$ is the fraction of likes expressed by $u$ on conspiracy-related content, and hence $-1\leq\sigma_{u}\leq 1$. According to the sign of their polarization, users are labelled as science ($\sigma_{u}<0$) or conspiracy ($\sigma_{u}>0$) supporters. 
Note that we here ignore the commenting activity since a comment may be an endorsement, a criticism, or even a response to a previous comment.

Furthermore, we define the engagement $\psi(u)$ of a user $u\in V^L$ as her liking activity normalized with respect to the number of likes of the most active user. Then, by defining $\theta(u)$ as the total number of likes expressed by $u$ on posts of $\mathcal{P}$, the following condition holds:
\begin{equation}
\psi(u)=\frac{\theta(u)}{\displaystyle\max_{v\in V^L}\theta(v)}.
\end{equation}

\subsection{The interaction network}\label{ssec:intnet}
In the analysis we investigate whether the mediated context of interaction of the comment thread leads to lexical convergence between co-commenters as their level of interaction grows. The most obvious metrics for quantifying this kind of interaction are given by the total number of posts co-commented and by the total number of comments. Nevertheless, these two measures may have some counter-intuitive limitations.
As an example, let us consider the user pair $\{u,v\}$ co-commenting on post $p_1,\ldots,p_{10}$ and the user pair $u^{\prime}, v^{\prime}$ co-commenting on post $p^{\prime}_1,\ldots,p^{\prime}_{5}$ according to the following tables:

\begin{table}[!ht]
\small\centering
\begin{tabular}{lcccccccccc}
	\toprule
	& $p_1$ & $p_2$ & $p_3$ & $p_4$ & $p_5$ & $p_6$ & $p_7$ & $p_8$ & $p_9$ & $p_{10}$\\
	\midrule
	$u$ & 3 & 3 & 3 & 3 & 3 & 3 & 3 & 3 & 3 & 3\\
	$v$ & 3 & 3 & 3 & 3 & 3 & 3 & 3 & 3 & 3 & 3\\
	\bottomrule
\end{tabular}
\end{table}
\begin{table}[!ht]
\small\centering
\begin{tabular}{lccccc}
\toprule
& $p^{\prime}_1$ & $p^{\prime}_2$ & $p^{\prime}_3$ & $p^{\prime}_4$ & $p^{\prime}_5$\\
\midrule
$u^{\prime}$ & 8 & 8 & 8 & 8 & 8\\
$v^{\prime}$ & 16 & 16 & 16 & 16 & 16\\
\bottomrule
\end{tabular}
\end{table}

\noindent Here, the cell values of each table correspond to the number of comments a given user makes on a given post. Notice that, since we are considering co-commenting, we are selecting posts for which both users produce at least one comment.

We observe that counting the number of co-commented posts can penalize pairs of users who have an high number of comments on few posts, while counting the total number of comments two users leave on the same posts does not take into account for unbalanced contributes of the two commenters. 

In fact, the number of co-commented posts is $10$ for the pair $u,v$ and $5$ for the pair $u^{\prime},v^{\prime}$, even if the first pair shares a lower number of comments ($60$) with respect to the second pair (that shares $120$ comments).
On the other hand, the total number of comments by $u^{\prime},v^{\prime}$ on each post is very high ($24$) even if the activity of $v^{\prime}$ is twice the activity of $u^{\prime}$.

To overcome these limitations, we define a metric that takes into account not only the number of different posts users interact with, but also the unbalance among users' comments on single posts. More formally, let $c_u(p)$ be the set of comments that user $u$ expressed on $p\in\mathcal{P}$ and let $\mathcal{P}_{uv}$ be the subset of $\mathcal{P}$ where both users $u$ and $v$ commented - i.e., $\mathcal{P}_{uv}=\{p\in\mathcal{P}\,\,\vert\,\, c_u(p)\neq\emptyset\,\text{ and }\,c_v(p)\neq\emptyset\}$. We define the interaction level between $u$ and $v$ as
\begin{equation}\label{eq:int_level}
I_{uv}=\displaystyle\sum_{p\in\mathcal{P}_{uv}}\min_p(\lvert c_u(p)\rvert, \lvert c_v(p)\rvert).
\end{equation}
We could think to this measure as a counter of interactions back-and-forth between $u$ and $v$.
By solving Eq. (\ref{eq:int_level}) for the user pairs of the previous example, we obtain $I_{uv}=30$ and $I_{u^{\prime}v^{\prime}}=40$, respectively, which represent more reasonable values for their amount of interaction.

We use Eq. (\ref{eq:int_level}) to weight the links of the interaction network with nodes representing science and conspiracy users, and a link existing between a pair of users if they co-commented at least once. More formally, let $G=\{\mathcal{C},\mathcal{P},E\}$ be the bipartite network whose vertex parts $\mathcal{C}$ and $\mathcal{P}$ denote the set of commenters and the set of posts in our collection, respectively. Here the link $\{u,p\}\in E$ between the user $u$ and the post $p$ exists if $u$ commented $p$. By projecting on the level of users we obtain the weighted interaction network $G_{\mathcal{C}}=\{\mathcal{C},E_I\}$ where the link $\{u,v\}\in E_I$ exists if both users $u$ and $v$ commented at least once on the same post $p$ and the link weight equals their interaction level $I_{uv}$. 

\subsection{Backbone detection algorithm}
The disparity filter algorithm is a network reduction technique that identifies the backbone structure of a weighted network without destroying its multiscale nature \cite{Serrano2009}. We use this algorithm to determine the connections that form the backbones of our interaction network and to produce clear visualization.

\subsection{Lexical convergence measurement}\label{ssec:bow}
To test whether more conversation leads to lexical convergence, we perform a pairwise comparative analysis on the multisets of words used by individuals who interact with each other through co-commenting activity. Specifically, we associate to each user her normalized bag of words (BOW) disregarding grammar and even word order but keeping multiplicity \cite{Zipf1949}. A user BOW is a sparse vector of occurrence counts of words - i.e, a sparse histogram over the dataset vocabulary.
We represent BOW objects as a Vector Space Model (VSM) \cite{Salton1988} with normalized term frequencies as components, thus, in the remainder of the paper, we say BOW for referring to normalized BOW.

To measure the lexical convergence $\ell_{uv}$ between two co-commenters $u$ and $v$, we calculate the normalized dot-product - i.e. the cosine similarity \cite{Salton1986}, of their BOWs. More formally, if $\mathbf{x}^u=(x^{u}_1,x^{u}_2,\dots,x^{u}_n)$ and $\mathbf{x}^v=(x^{v}_1,x^{v}_2,\dots,x^{v}_n)$ are the BOWs of $u$ and $v$, respectively, their lexical convergence is computed as follows:
\begin{equation}\label{eq:cosine}
\ell_{uv}=\frac{\mathbf{x}^u\cdot\mathbf{x}^v}{\lVert\mathbf{x}^u\rVert\lVert\mathbf{x}^v\rVert}=\frac{\displaystyle\sum_{k=1}^nx^{u}_k x^{v}_k}{\sqrt{\displaystyle\sum_{k=1}^n(x^{u}_k)^2}\sqrt{\displaystyle\sum_{k=1}^n(x^{v}_k)^2}}
\end{equation}
Since all the components of a BOW have non negative values, it follows that the cosine similarity between two any BOWs, hence the lexical convergence between the corresponding users, varies in the range $[0,1]$.

\subsection{Spearman's rank correlation coefficient}\label{ssec:spe}
Spearman's rank correlation coefficient $r_s$ is a measure of the statistical dependence between the rankings of two variables \cite{Spearman1904}. It assesses how well the relationship between two variables can be described using a monotonic function, by returning a value from $-1$ to $1$, where $1$ means a perfect positive correlation between ranks, $-1$ a perfect negative correlation between ranks, and $0$ no correlation between ranks.
The Spearman's $r_s$ is the nonparametric version of the Pearson's correlation coefficient. Namely, for two variables $x$ and $y$, it is
\begin{equation}\label{eq:spearman}
r_s(x,y)=\frac{\text{cov}\left(\text{rg}_x,\text{rg}_y\right)}{\sigma_{\text{rg}_x}\sigma_{\text{rg}_y}}
\end{equation}
where $\text{cov}\left(\text{rg}_x,\text{rg}_y\right)$ denotes the covariance of the rank variables, and $\sigma_{\text{rg}_x}$ and $\sigma_{\text{rg}_y}$ are the standard deviations of the rank variables.

We use $r_s$ for testing whether the lexical convergence measurements of a pair of co-commenters exhibits a monotonically increasing pattern over time. Moreover, in order to determine significance we perform a Student's $t$-test by calculating
\begin{equation}
t=r_s\sqrt{\frac{n-2}{1-r_s^2}}
\end{equation}
which is distributed approximately as Student's $t$ distribution with $n-2$ degrees of freedom under the null hypothesis \cite{Siegel1988}.

\section{Results and discussion}
\subsection{Quantifying user polarization}
Our analysis starts by investigating the polarization of users towards science and conspiracy. Users are labelled by means of a simple thresholding algorithm accounting for the percentage of likes on one or the other category. Formally, the polarization of a user $u\in V^L$ is a real number $\sigma_{u}\in[-1,1]$ and $u$ is labelled as science or conspiracy user according to whether $\sigma_{u}<0$ or $\sigma_{u}>0$, respectively (See \nameref{sec:mm} for further details). 

\begin{figure}[!ht]
\centering
\includegraphics[width=0.5\linewidth]{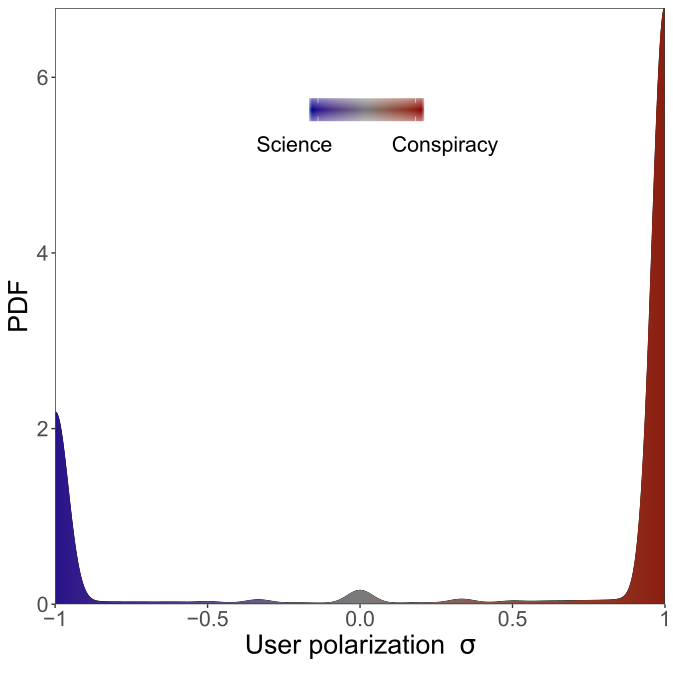}
\caption{\footnotesize{\bf Probability density function (PDF) of users' polarization.} Notice the strong bimodality of the distribution, with two sharp peaks localized at $-1\lesssim\sigma\lesssim-0.95$ (science users) and at $0.95\lesssim\sigma\lesssim1$ (conspiracy users).}
\label{fig:userpol}
\end{figure}

The probability density function (PDF) of users' polarization shows that users are clearly split into two groups with opposing polarization, whereas few users are weakly polarized or unpolarized (Figure \ref{fig:userpol}). To better define the properties of these groups, we consider the set $V^L_{\text{science}}$ of users with polarization more than $95\%$ towards science
\[
V^L_{\text{science}} = \{u\in V^L;\,\sigma(u)\leq-0.95\},
\]
and the set $V^L_{\text{conspir}}$ of users with polarization more than $95\%$ towards conspiracy
\[
V^L_{\text{conspir}} = \{u\in V^L;\,\sigma(u)\geq0.95\};
\]
such sets correspond to the two peaks of the bimodal distribution of Figure \ref{fig:userpol} and their cardinalities show that most of the users are highly polarized: $\lvert V^L_{\text{science}}\rvert=243,977$ and $\lvert V^L_{\text{conspir}}\rvert=758,673$. This confirms the existence of two well-formed and highly segregated communities around scientific and conspiracy topics \cite{Bessi2014,BessiPONE1-2015,Bessi2015,Bessi2016,Brugnoli2019,ZolloPONE2015} - i.e., users are mainly active in only one category.

\begin{figure}[!ht]
\centering
\includegraphics[width=\linewidth]{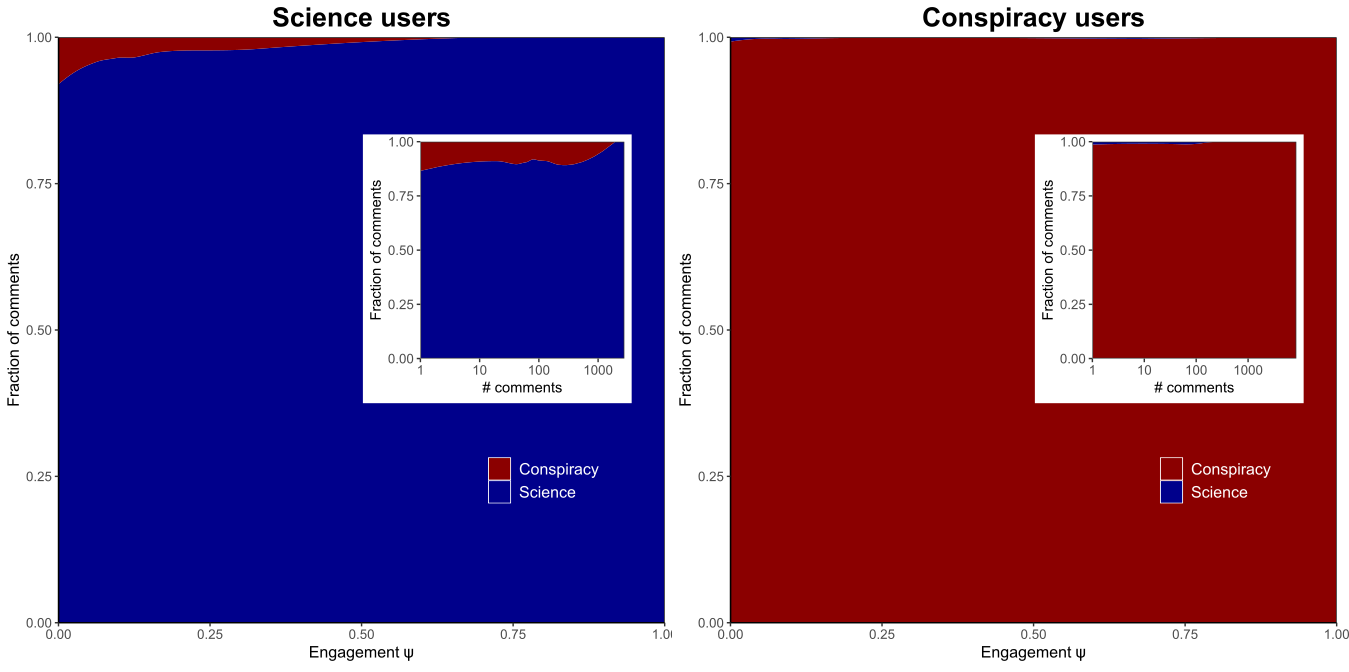}
\caption{\footnotesize{\bf Fraction of comments as a function of the engagement and the number of comments, respectively.} Left panel: for a polarized user $u\in V^L_{\text{science}}$, the fraction of comments on contents produced by science pages is $\gtrsim 0.93$ and grows with the engagement $\psi$. Right panel: almost the entire commenting activity ($\gtrsim 99\%$) of a polarized user $u\in V^L_{\text{conspir}}$ is on contents produced by conspiracy pages, and it is actually a total as the engagement $\psi$ grows. Inset plots show similar trends for the same quantities as a function of the number of comments.}
\label{fig:comments}
\end{figure}

As a further step, we focus on the commenting activity of polarized users. Far from promoting shared interests between the co-commenters, science and conspiracy pages could be co-opted as contexts in which rapport-threatening disagreements take place. Nevertheless, we find that most of the users of $V^L_{\text{science}}$ ($V^L_{\text{conspir}}$) almost exclusively comment on just science (conspiracy) pages. Figure \ref{fig:comments} shows the fraction of comments of polarized users as a function of their engagement $\psi(\cdot)$ both in the case of users polarized toward science (left panel) and in the case of users polarized toward conspiracy (right panel). Inset plots show the same quantities as a function of the number of comments they wrote.

Hence, posts where polarized users of both communities commented are extremely rare and mainly confined on the conspiracy side ($5,760$ on a total of $7,751$). Users polarized towards science or conspiracy mainly concentrate their commenting activity on posts from pages of their own community, and this trend grows both with the engagement and with the number of comments expressed.

\subsection{Co-commenting interactions}\label{ssec:cocomm}
\begin{figure}[!ht]
\centering
\includegraphics[width=0.5\linewidth]{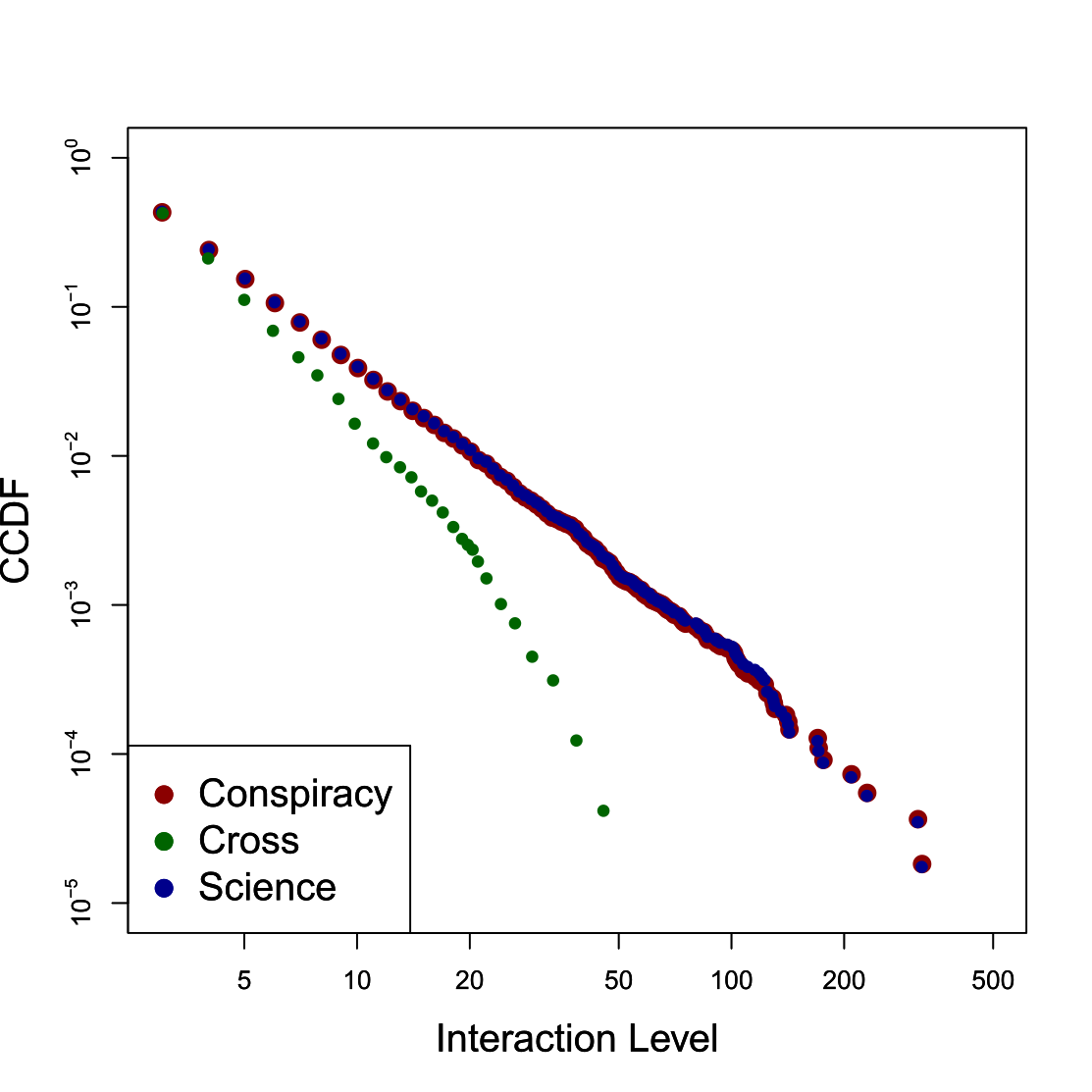}
\caption{\footnotesize{\bf Interaction level.}
Empirical complementary cumulative distribution function (CCDF) of users' interaction level broken down by interaction type. All the distributions are heavy–tailed.}
\label{fig:intlev}
\end{figure}

Comments represent the way in which collective debates take form around the topic promoted by posts. In order to measure the amount of interaction between two co-commenters $u$ and $v$, we consider the interaction level $I_{uv}$ given by Eq. (\ref{eq:int_level}).
Consequently, we investigate the overall distribution of the interaction level across each pair of co-commenters in $G_{\mathcal{C}}$, looking at the extent to which certain pairs possess more activity than others. For this, we calculate the empirical complementary cumulative distribution function (CCDF) of the interaction level $I_{uv}$ broken down by interaction type, as represented in a double logarithmic axes plot in Figure \ref{fig:intlev}. The plot shows that all the three distributions are heavy–tailed (possibly power-law) - i.e. the majority of the user pairs display very little interaction, while only few users are highly interacting in the sense of Eq. (\ref{eq:int_level}).

\begin{figure}[!ht]
\centering
\includegraphics[width=0.5\linewidth]{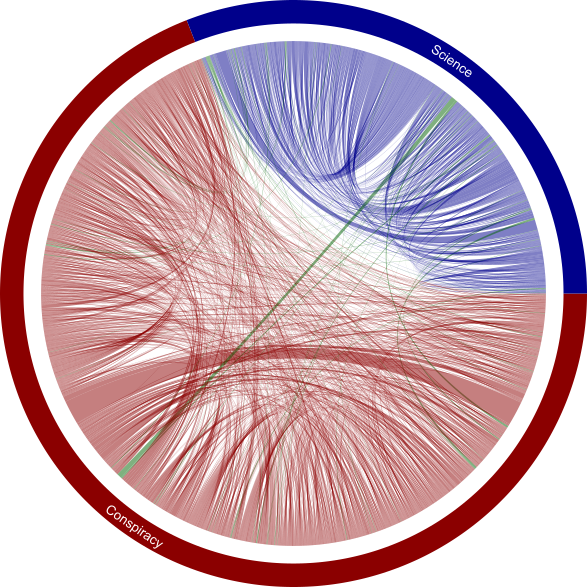}
\caption{\footnotesize{\bf The network of interactions.}
Backbone of the interaction network $\overline{G}_{I}^{\mathcal{C}}$. Link colors indicate the interaction type: red for interactions between conspiracy users, green for cross interactions and blue for interactions between science users. The thickness of a line indicates the strength of the link, i.e. the interaction level between the nodes connected by the link. Nodes are ordered according to the community membership. Users polarized towards science or conspiracy mainly concentrate their commenting activity on posts from pages of their own community, with very few cross interactions.}
\label{fig:graph}
\end{figure}

Therefore, in order to further analyze the process of lexical convergence, we focus only on pairs of polarized users exhibiting a reasonably significant interaction level instead of considering the interaction network as a whole.
Namely, hereafter, we write interaction network for referring to the induced subgraph $\overline{G}^L_{\mathcal{C}}$ of $G_{\mathcal{C}}=\{\mathcal{C},E_I\}$ whose vertices are all the commenters in $V^L_{\text{science}}\cup V^L_{\text{conspir}}$ and whose edge set consists of all the edges $\{u,v,\}\in E_I$ with both endpoints in $V^L_{\text{science}}\cup V^L_{\text{conspir}}$ such that $I_{uv}\geq3$ (See \nameref{sec:mm} for further details).
The interaction network $\overline{G}_{\mathcal{C}}^L$ contains 15,034 users and 57,664 links, and it is composed by 11,949 conspiracy users and 3,085 science users who produced 46,153 and 10,998 homophilic interactions, respectively. The cross interactions are 513, generated by 474 users (242 from conspiracy and 232 from science) who commented posts from pages of the other community.
Figure \ref{fig:graph} shows the backbone of $\overline{G}_{I}^{\mathcal{C}}$. The thickness of a line indicates the strength of the link, i.e. the interaction level among the nodes connected by the link. 

\subsection{Word-level discrimination between science and conspiracy languages}
Before exploring differences between the language used by users polarized towards science with the language used by users polarized towards conspiracy, we cleaned the corpora of their comments. Namely, we first removed URLs and emails, then we performed sentence segmentation, tokenization, POS tagging and lemmatization by means of UDPipe 2.0 pipeline \cite{udpipe2017}. The model was trained both on the Italian Stanford Dependency Treebank \cite{Bosco2013} and on a collection of social media texts \cite{Sanguinetti2018}, and only words tagged as noun, proper noun, adjective, verb (except auxiliaries and modals), adverb of negation (\textit{no, not, nor, neither,\dots}) were retained. All words were converted to lower case and reduced into their lemma form (for instance, \textit{goes, went, gone} reduced into \textit{go}). The vocabulary of the pre-processed dataset consists of 83,615 unique terms, whereas the words used by science and conspiracy users are 56,743 and 62,830, respectively. Moreover, science and conspiracy users share 38,258 words, which, counted with their multiplicity, represent $\gtrsim97\%$ of the total words used in both the communities. Hence, even if the two communities have vocabularies that differ by $\sim 30\%$ of their words, their users resort mostly on the remaining $\sim 70\%$ common words for commenting their preferred content.

\begin{figure}[!ht]
\centering
\includegraphics[width=0.5\linewidth]{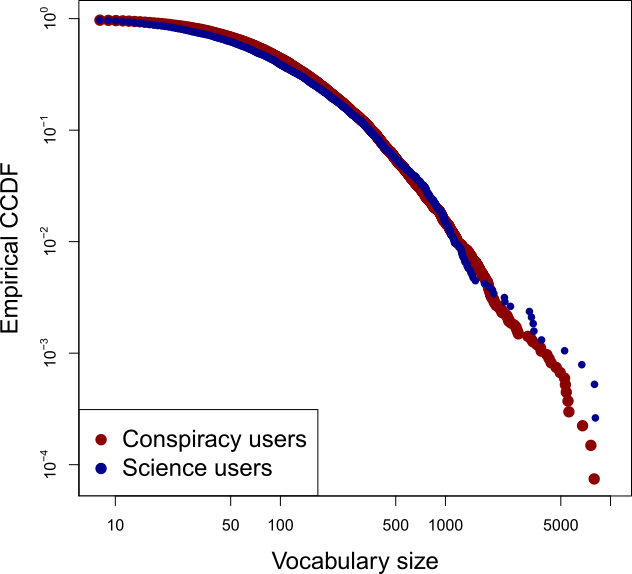}
\caption{\footnotesize{\bf User vocabulary size.}
Empirical complementary cumulative distribution function (CCDF) of science and conspiracy users' vocabulary size. Both the distributions are indicating similar patterns.}
\label{fig:vocsiz}
\end{figure}

Figure \ref{fig:vocsiz} shows a double logarithmic axes plot of the CCDF of science and conspiracy users' vocabulary size, respectively. Both distributions are indicating a similar pattern. Median and mean vocabulary size of science users are $\sim73$ and $\sim160$ words, respectively. Similarly, median and mean vocabulary size of conspiracy users are $\sim88$ and $\sim166$ words, respectively.

\begin{figure}[!ht]
\centering
\includegraphics[width=\linewidth]{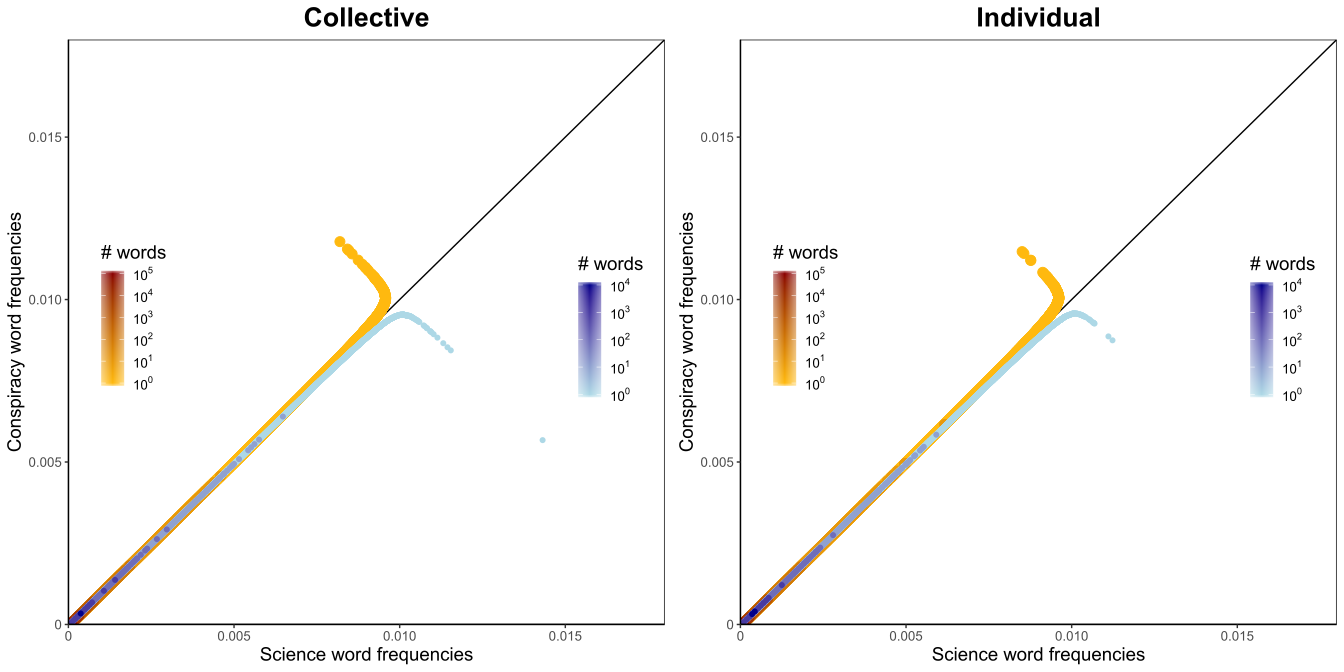}
\caption{\footnotesize{\bf Comparison of word frequencies between science and conspiracy users.}
Left panel: Collective level. Right panel: Individual level. Points that are close to the bisector represent words having similar frequencies in both the communities. Points that are far from the bisector represent words that are found more in one community than another. Words are grouped (and ordered) by the absolute difference of frequency from one community to another, and the distance point-line equals such difference relative to the words represented by the point. Points are colored according to the number of words exhibiting the corresponding frequency difference in favour of conspiracy (red gradient) or science (blue gradient).}
\label{fig:worduse}
\end{figure}

We also analyze the frequency of words appearing in science ($f_s$) and conspiracy ($f_c$) comments, respectively, both at collective and individual levels. For each word, in the former case, we compare the frequency of usage in both the communities of users while, in the latter case, we consider the fraction of science and conspiracy commenters who used that word. Points of Figure \ref{fig:worduse} that are close to the line represent words having similar frequencies in both the communities. Points that are far from the line represent words that are found more in one community than another. Words are grouped (and ordered) by the absolute difference of frequency from one community to another, and the distance point-line equals such difference relative to the words represented by the point. Moreover, points are colored according to the number of words exhibiting the corresponding frequency difference in favour of conspiracy (red gradient) or science (blue gradient).

Plots clearly show that most of the words have a similar frequency usage both by the overall communities and by their individual members.
Remarkable differences are limited to a minority of words, which in turn provide some insights about the kind of information processed by the two communities of users and about the sentiment expressed in their comments. Tables \ref{table2} and \ref{table3} suggest that science users discuss more on subjects related to nature and scientific research, whereas conspiracy users are more prone to discuss of politics and economy.

\begin{table}[!ht]
\small\centering
\caption{The most distinctive words of science community\label{table2}}
\begin{tabular}{lc|lc}
\toprule
Collective & $f_s-f_c$ & Individual & $f_s-f_c$\\
& ($\times10^{-3}$) & & ($\times10^{-3}$)\\
\midrule
\textit{animale} & 6.1 & \textit{animale} & 1.7\\
\textit{scienza} & 2.2 & \textit{scienza} & 1.6\\
\textit{sperimentazione} & 2.0 & \textit{sperimentazione} & 1.0\\
\textit{ricerca} & 1.9 & \textit{ricerca} & 1.0\\
\textit{medico} & 1.6 & \textit{medico} & 1.0\\
\textit{specie} & 1.5 & \textit{scientifico} & 0.9\\
\textit{scientifico} & 1.4 & \textit{animalista} & 0.8\\
\textit{vaccino} & 1.3 & \textit{specie} & 0.8\\
\textit{animalista} & 1.3 & \textit{natura} & 0.8\\
\textit{studio} & 1.2 & \textit{fantastico} & 0.7\\
\bottomrule
\end{tabular}
\end{table}

\begin{table}[!ht]
\small\centering
\caption{The most distinctive words of conspiracy community\label{table3}}
\begin{tabular}{lc|lc}
\toprule
Collective & $f_c-f_s$ & Individual & $f_c-f_s$\\
& ($\times10^{-3}$) & & ($\times10^{-3}$)\\
\midrule
\textit{popolo} & 2.6 & \textit{Italia} & 2.1\\
\textit{politica} & 2.3 & \textit{politica} & 2.1\\
\textit{Italia} & 2.2 & \textit{italiano} & 1.8\\
\textit{italiano} & 2.0 & \textit{popolo} & 1.8\\
\textit{guerra} & 1.8 & \textit{governo} & 1.2\\
\textit{mondo} & 1.6 & \textit{paese} & 1.2\\
\textit{Dio} & 1.5 & \textit{guerra} & 1.2\\
\textit{governo} & 1.5 & \textit{schifo} & 1.1\\
\textit{potere} & 1.4 & \textit{pagare} & 1.1\\
\textit{scia} & 1.3 & \textit{soldi} & 1.1\\
\bottomrule
\end{tabular}
\end{table}

This is also confirmed by analyzing the words used in just one of the two communities (Tables \ref{table4} and \ref{table5}). Words only from science consist almost exclusively of technical terms, acronyms and names of people linked to biology and scientific research (mainly about evidence-based medicine, pharmacology and animal testing). Instead, words only from conspiracy are mainly related to politics and economy, religion and healthcare (mainly about alternative diet). Moreover, this latter word set contains a large group of terms, mainly verbs, which clearly describe a negative sentiment toward strong powers as well as a community spirit to fight back: e.g. \textit{unirci, aguzzino, ubbidire, sommossa, purificare, governarci, armiamoci, dimettiti, ladrocinio, potentato}. See Supplemental material for more exhaustive word lists.

\begin{table}[!ht]
\small\centering
\caption{The most frequent words used solely by science users\label{table4}}
\begin{tabular}{lc|lc}
\toprule
Collective & $f_s$ & Individual & $f_s$\\
& ($\times10^{-5}$) & & ($\times10^{-5}$)\\
\midrule
\textit{stabulario} & 5.5 & \textit{istrice} & 4.7\\
\textit{thalidomide} & 5.0 & \textit{stabulario} & 4.2\\
\textit{Grignaschi} & 4.5 & \textit{lanice} & 3.5\\
\textit{lanice} & 4.2 & \textit{esapode} & 3.4\\
\textit{esapode} & 4.1 & \textit{Grignaschi} & 3.2\\
\textit{istrice} & 3.8 & \textit{drosophila} & 3.0\\
\textit{ecvam} & 3.5 & \textit{natrix} & 3.0\\
\textit{antivivisezionista} & 3.2 & \textit{mantis} & 2.7\\
\textit{mesenchimali} & 3.1 & \textit{terrario} & 2.7\\
\textit{natrix} & 3.1 & \textit{anellide} & 2.5\\
\bottomrule
\end{tabular}
\end{table}

\begin{table}[!ht]
\small\centering
\caption{The most frequent words used solely by conspiracy users\label{table5}}
\begin{tabular}{lc|lc}
\toprule
Collective & $f_c$ & Individual & $f_c$\\
& ($\times10^{-5}$) & & ($\times10^{-5}$)\\
\midrule
\textit{satanismo} & 6.7 & \textit{satanismo} & 5.5\\
\textit{stevia} & 5.5 & \textit{Eurogendfor} & 5.2\\
\textit{Eurogendfor} & 5.2 & \textit{stevia} & 5.2\\
\textit{Bonino} & 4.7 & \textit{Bonino} & 4.6\\
\textit{yogurt} & 4.5 & \textit{yogurt} & 4.5\\
\textit{gni} & 4.3 & \textit{zenzero} & 4.1\\
\textit{Ighina} & 4.2 & \textit{svegliarci} & 3.8\\
\textit{zenzero} & 3.5 & \textit{cancelliere} & 3.8\\
\textit{cancelliere} & 3.4 & \textit{scioperare} & 3.3\\
\textit{Talmud} & 3.3 & \textit{Ighina} & 3.2\\
\bottomrule
\end{tabular}
\end{table}

\subsection{The lexical evolution of polarized co-commenters}
Here we want to test whether more conversation (intended as co-commenting activity) leads to lexical convergence. Namely, we want to investigate whether the lexical choices of co-commenters represent a proxy for the emergence of polarization and collective identities.

\begin{figure}[!ht]
\centering
\includegraphics[width=\linewidth]{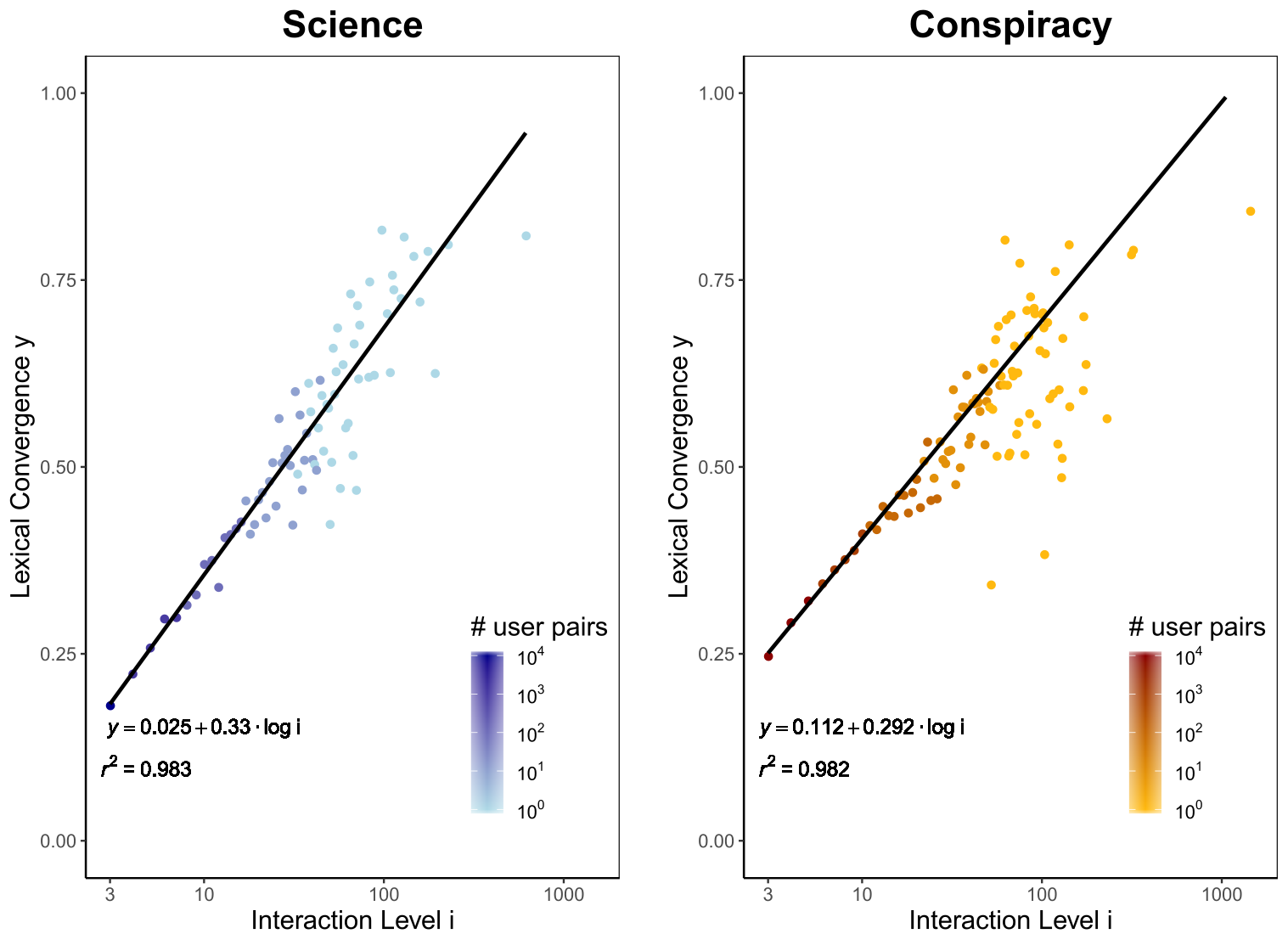}
\caption{\footnotesize{\bf Lexical convergence as a function of the level of interaction.} Left panel: interactions between users both polarized towards science. Right panel: interactions between users both polarized towards conspiracy. In both the panels, we plot the average lexical convergence between all the co-commenter pairs $\{u,v\}$ such that $I_{uv}=i$, versus the interaction level $\log i$. Points are colored according to the number of pairs of co-commenters who reached the corresponding interaction level. Full lines are the results of a linear regression model $y(u,v)=\beta_0+\beta_1\log(i)$. Coefficients are estimated using weighted least squares with weights given by the total number of pairs of co-commenters per interaction level. All the p-values are close to zero.}
\label{fig:Cos_Int}
\end{figure}

First, we compare the BOWs of more than 57K pairs of interacting users (representing the links of $\overline{G}_{I}^{\mathcal{C}}$) who made $\sim$428K comments in a time span of 5 years (see \nameref{sec:mm}).
For each pair of co-commenters $\{u,v\}$, Figure \ref{fig:Cos_Int} shows the lin-log plot of the average lexical convergence, respect given interaction level $I_{uv}$, for interactions between science users (left panel) and interactions between conspiracy users (right panel). Points are colored according to the number of pairs of co-commenters who reached the corresponding interaction level.

Plots suggests in both the cases a linear correlation between the variables; thus, we check whether for a given interaction level $i$, the average lexical convergence $y(u,v)$ between all the co-commenter pairs $\{u,v\}$ such that $I_{uv}=i$, can be predicted by means of a linear regression model where the explanatory variable is a logarithmic transformation of the interaction level $i$, i.e. $y(u,v)=\beta_0+\beta_1\log(i)$. Coefficients are estimated using weighted least squares with weights given by the total number of pairs of co-commenters per interaction level and they are – with the corresponding standard errors inside the round brackets – $\beta_0=0.025\, (0.003)$ and $\beta_1=0.330\, (0.005)$, with $r^2=0.983$, for interactions between science users; $\beta_0=0.112\, (0.002)$ and $\beta_1=0.292\, (0.004)$, with $r^2=0.982$, for interactions between conspiracy users. All the p-values are close to zero. This suggests the existence of positive correlation between interaction level and lexical convergence inside echo chambers.

\begin{figure}[!ht]
\centering
\includegraphics[width=\linewidth]{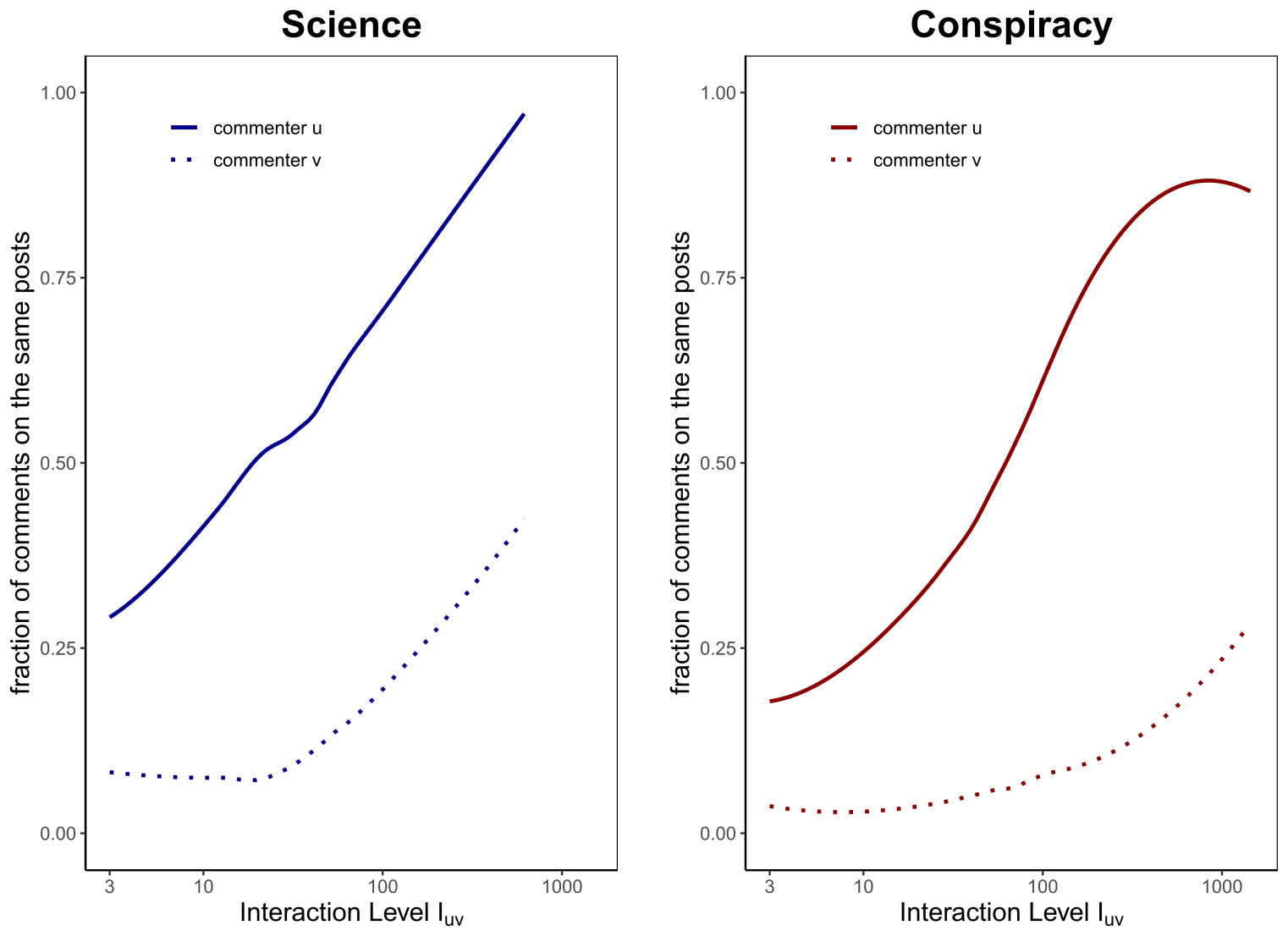}
\caption{\footnotesize{\bf Fraction of comments expressed by co-commenters on the same posts as a function of their interaction level.} At least for one of the two co-commenters, the number of comments expressed on the same posts represent a moderate fraction of the total number of her comments. This ensures that the lexical convergence observed is not topic-dependent.
}
\label{fig:comm_perc}
\end{figure}

One could argue that such a trend might be primarily due to the fact that, in comment threads, commenters respond more likely to the post than to other comments. Hence the lexical choices would be merely dictated by the content promoted by the post. However, for comparing the lexicon used by co-commenters, we consider their entire vocabularies and not only those words used in the co-commented posts. Figure \ref{fig:comm_perc} shows the average fraction of comments expressed by the pairs of co-commenters $\{u,v\}$ on the same posts respect given interaction level $I_{uv}$.

Plots clearly show that, at least for one of the two co-commenters, the number of comments expressed on the same posts represent a moderate fraction of the total. This ensures that the observed lexical convergence phenomenon is not topic-dependent. 

Nevertheless, by considering the entire vocabularies of co-commenters, the results of lexical convergence might merely indicate that when users comment more, they tend to multiply commonly used words, therefore increasing the proportion of words in common with the other users who comment a lot. For cross-validating our outcomes, we performed a randomization test \cite{Noreen1989} generating 1000 instances of the actual dataset broken down by interaction type (pseudosamples). Specifically, for both the types of interaction, pseudosamples are generated by randomly permuting the users' BOWs and by repeating the lexical convergence measurements. The permutation test shows that the observed data are statistically significant with a p-value $<10^{-3}$.
 
Furthermore, for each link $\{u,v\}\in\overline{G}_{I}^{\mathcal{C}}$ with weight $I_{uv}\geq 8$, we measure how lexical convergence between its endpoints $u$ and $v$ evolves over time. 
This setting ensures an examination of the lexical convergence evolution at least through six consecutive interactions.
Indeed, to make our analysis not dependent on the physical time, we measure the time when $u$ and $v$ co-commented in terms of the interaction level $I_{uv}$; moreover, to be consistent with the analysis of Figure \ref{fig:Cos_Int}, we start measuring lexical convergence when $u$ and $v$ currently co-commented 3 times.

\begin{figure}[!ht]
\centering
\includegraphics[width=\linewidth]{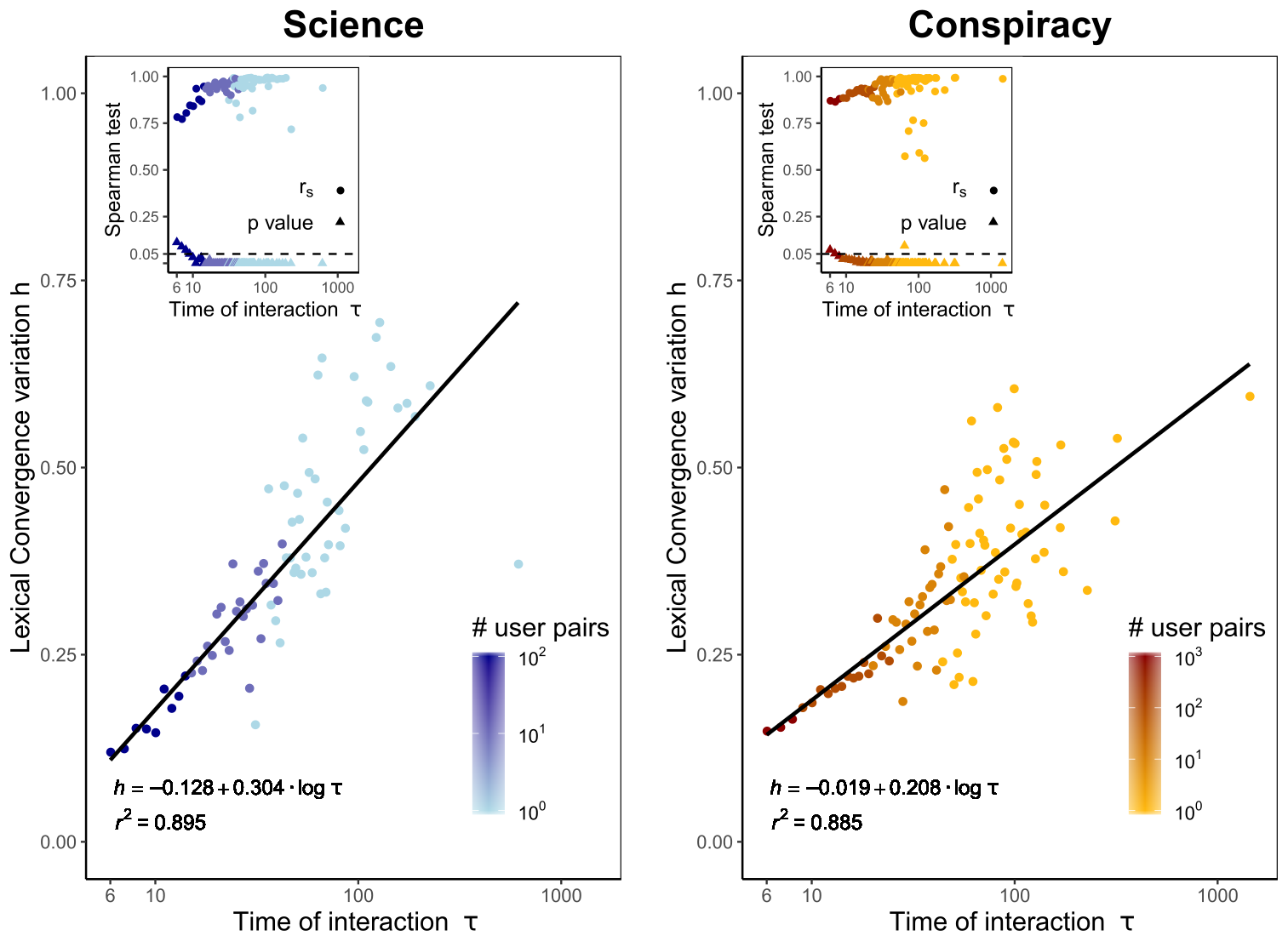}
\caption{\footnotesize{\bf Lexical convergence evolution through time.}
Left panel: interactions between users both polarized towards science. Right panel: interactions between users both polarized towards conspiracy.
Main plots show the mean lexical convergence increment $h$ for all the pairs of co-commenters per time of interaction $\tau$. Inset plots show the mean Spearman's rank correlation coefficient $r_s$ and the corresponding p-value with respect to the same variables. Points are colored according to the number of pairs of co-commenters active at the corresponding discrete-time point. Full lines are the results of a linear regression model $h(u,v)=\beta_0+\beta_1\log(\tau)$. Coefficients are estimated using weighted least squares with weights given by the total number of pairs of co-commenters active at every discrete-time point. All the p-values are close to zero.}
\label{fig:Time_Int}
\end{figure}

Let $\ell_{uv}(t)$ be the measurement of lexical convergence between the co-commenters $u$ and $v$ at time $t$. Moreover, let $\tau_{\min}(u,v)=3$ and $\tau_{\max}(u,v)=I_{uv}$ be the first and the last discrete-time point when $u$ and $v$ co-commented, respectively. Then $\tau(u,v)=\tau_{\max}(u,v)-\tau_{\min}(u,v)$ denotes the length of time during which $u$ and $v$ interact through co-commenting activity. The main plots of Figure \ref{fig:Time_Int} show the mean variation $h(u,v)=\ell_{uv}(\tau_{\max})-\ell_{uv}(\tau_{\min})$ for all the pairs of co-commenters $\{u,v\}$ per time of interaction $\tau$, broken down by interaction type. The inset plots show the mean Spearman's rank correlation coefficient $r_s$ and the corresponding p-value with respect to the same variables (see \nameref{sec:mm}).

Both panels (left: both co-commenters polarized towards science, right: both co-commenters polarized towards conspiracy) suggest a linear correlation between the variables; thus, we check whether for all the pairs $\{u,v\}$ who co-commented for a period of length $\tau$, the average variation of lexical convergence $h(u,v)$ can be predicted by means of a linear regression model where the explanatory variable is a logarithmic transformation of $\tau$, i.e. $h(u,v)=\beta_0+\beta_1\log(\tau)$. Coefficients are estimated using weighted least squares with weights given by the total number of pairs of co-commenters per length of co-commenting period and they are – with the corresponding standard errors inside the round brackets – $\beta_0=-0.128\, (0.013)$ and $\beta_1=0.304\, (0.012)$, with $r^2=0.895$, for interactions between science users; $\beta_0=-0.019\, (0.008)$ and $\beta_1=0.208\, (0.008)$, with $r^2=0.885$, for interactions between conspiracy users. All the p-values are close to zero. Moreover the corresponding inset plots indicate that the overall lexical convergence increments of the main plots follow a monotonically increasing pattern over time.

As stressed in \nameref{ssec:cocomm}, the cross interactions represent a very small fraction of the total pairs of co-commenters who exhibit a reasonably significat interaction level.
With the additional constraint $I_{uv}\geq8$, the number of cross links $\{u,v\}\in\overline{G}_{I}^{\mathcal{C}}$ reduces to only $22$, thus not guaranteeing adeguate statistical power to the analysis. However, for such pairs of co-commenters, the lexical convergence increments show a much broader dispersion ($r^2=0.287$) around the line of best fit which in turn exhibits a smaller slope ($0.133$) than the regressions of Figure \ref{fig:Time_Int}.

\section{Conclusion}
In this work we analyze the lexicons used by the communities of users emerging on Facebook around verified and unverified contents.

Despite the high level of segregation between the two communities of users, both in terms of liking activity and commenting activity, we show that most of the words are used with a similar frequency both at collective and individual level. Nevertheless, the minority of words exhibiting significant differences of frequency occurrence, provide important insights about the kind of information processed by the two communities of users and about the sentiment expressed in their comments. 

Moreover, by focusing on the comment thread of a post, we test whether more conversation (intended as co-commenting activity) between polarized users leads to lexical convergence. Our findings reveal a strong positive correlation between the lexical convergence of co-commenters and their number of interactions. This suggests that such a trend could be a proxy for the emergence of collective identities and polarization in opinion dynamics.
Nonetheless, the fact that even users with opposing views try to coordinate their lexical choices when interact with the same posts, suggests that a dialogue between competing parties is possible and indeed it should be stimulated in an attempt to smooth polarization and to reduce both the risk and the consequences of misinformation. 

We emphasize that our investigation is restricted to lexicon. Future works will be devoted to perform a semantic analysis and a comparison of the comment corpora generated by the two communities of users.

\end{document}